# Comet 29P/Schwassmann-Wachmann 1 dust environment from photometric observation at the SOAR Telescope

Enos Picazzio[*,a], Igor V. Luk'yanyk[b], Oleksandra V. Ivanova[c,d], Evgenij Zubko[e], Oscar Cavichia[f], Gorden Videen[g,h], Sergei M. Andrievsky[i,j]

[a] *Instituto de Astronomia, Geofísica e Ciências Atmosféricas, Departamento de Astronomia, Rua do Matão, 1226, Cidade Universitária*
[b] *Astronomical Observatory, Taras Shevchenko National University of Kyiv, Observatorna str. 3, Kyiv 04053, Ukraine*
[c] *Astronomical Institute of the Slovak Academy of Sciences, Tatranská Lomnica, SK-05960, Slovak Republic*
[d] *Main Astronomical Observatory, National Academy of Sciences of Ukraine, Goloseevo, Kyiv 03680, Ukraine*
[e] *School of Natural Sciences, Far Eastern Federal University, 8 Sukhanova Street, Vladivostok 690950, Russia*
[f] *Instituto de Física e Química, Universidade Federal de Itajubá, Av. BPS, 1303, Pinheirinho, Itajubá CEP 37500-903, Brasil*
[g] *US Army Research Laboratory, RDRL-CIE-S, 2800 Powder Mill Road, Adelphi MD 20783, USA*
[h] *Space Science Institute, 4750 Walnut Street, Boulder Suite 205, CO 80301, USA*
[i] *Department of Astronomy and Astronomical Observatory, Odessa National University, Shevchenko Park, Odessa 65014, Ukraine*
[j] *GEPI, Observatoire de Paris, PSL Research University, CNRS, Place Jules Janssen, Meudon 92195, France*

## ABSTRACT

We report photometric observations of comet 29P/Schwassmann-Wachmann 1 made on August 12, 2016 with the broad-band B, V, R and I filters and the SOAR 4.1-meter telescope (Chile). We find the comet active at that time. Enhanced images obtained in all filters reveal three radial features in the 29P/Schwassmann-Wachmann 1 coma, regardless of the image-processing algorithm. Using a high-resolution spectrum of comet 29P/Schwassmann-Wachmann 1 reported by Ivanova et al. (2018) on the same date, we estimate the relative contribution of the gaseous emission and the continuum to the total response measured with our broadband B and V filters. The gaseous-emission contribution appears to be very small, 2.5%. We compute the dust production $Af\rho$ in 29P/Schwassmann-Wachmann 1 for the four filters and find its growth with the wavelength, from $3,393 \pm 93$ cm in the B filter to $8,561 \pm 236$ cm in the I filter. We model the color slope of dust in Comet 29P/Schwassmann-Wachmann 1 using agglomerated debris particles. Simultaneous analysis of the color slope in the B–R and R–I pairs suggests a single dominant chemical species of 29P/Schwassmann-Wachmann 1 dust particles consisting of Fe–Mg silicates and obeying a power-law size distribution with index $n \approx 2.55$. This conclusion is consistent with the previous thermal-emission study of 29P/Schwassmann-Wachmann 1.

### Introduction

Cometary dust includes remnants of primordial material from the zone of comet formation in addition to processed dust. In this sense, the centaur Comet 29P/Schwassmann-Wachmann 1 (hereafter 29P) is unique. Although this comet belongs to the Jupiter family of comets, an investigation based on an analysis of the dynamical evolution of the comet (Neslušan et al., 2017) based on orbital clones implies that the comet might have come to the planetary region from the Oort cloud, although its low orbital inclination does favour its origin in the trans-Neptunian belt. 29P presumably changed during its residence in the inner Solar System and its surface can no longer be regarded as primordial. Analysis of Spitzer observations (Stansberry et al., 2004; Kelley and Wooden, 2009) also shows that 29P is one out of many known comets, which contains a material that formed in a region hotter than that where cometary nuclei usually form. Dynamical evidence suggests that all Jupiter-family comets came to the inner Solar System from the trans-Neptunian region (e.g. Duncan et al., 2004), but the precise origin location is unknown. Dynamical and compositional studies of this comet are important, because they address this problem (A'Hearn, 2012). In addition the chemical composition investigation may help to understand why 29P, a comet in a nearly circular orbit at approximately 6 AU, beyond the sublimation zone of water, is continuously active and exhibits frequent bursts.

In this work, we analyze the color of 29P in order to obtain information about the chemical composition of its coma. We managed to organize observations of comet 29P to a sole epoch, 2016-08-12. As a consequence, our conclusion on chemical composition of its coma must be considered only with regard to this epoch. It is known that the color in comets could be the subject of fast and dramatic variations (e.g., Ivanova, 2017). This feature is still poorly addressed in the literature. Observations of comets are often time limited, and, as a consequence, observational data correspond typically to rare and random epochs. One of the most systematic surveys spanning a time period of nearly 5 year was conducted on C/1995 O1 (Hale-Bopp) (Weiler et al., 2003); the analysis of its color revealed short-term variations with a scale of a few days. These variations can be seen, for instance, on the bottom panel in Fig. 3 of Weiler et al. (2003) at the heliocentric distances of −4.6 AU, −2.86 AU, and +3.78 AU. Note that although this phenomenon can be clearly seen in Fig. 3 of the paper by Weiler et al. (2003), it did not draw much attention of the authors. Interestingly, such variations did not correspond with the heliocentric distance of the comet. A recent study of Comet C/2013 A1 (Siding Spring) with the Hubble Space Telescope also demonstrated that the color slope could vary by a factor of about two over a period of a few months (Li et al., 2014). Nevertheless, unlike in the case of Comet Hale-Bopp, the color of Comet Siding Spring remained red throughout all three epochs of observation.

To date, we accomplished quantitative analysis of various comets (e.g., Zubko 2011; 2014; 2015; 2016; Ivanova, 2017). Prior to this work, we found a chemically homogeneous coma only in one comet, 17P/Holmes (Zubko, 2011); whereas, in all other cases, the phase curves are consistent with at least two different refractory components of the coma.

### Observation and reduction

Photometric observations of comet 29P were made with Goodman Imaging/Spectrograph on SOAR 4.1-m telescope (Cerro Pachón, Chile) during August 12, 2016. The heliocentric and geocentric distances of the comet was r = 5.913 AU and Δ = 5.016 AU, respectively (see Table 1). The phase angle of the comet was 5.1°. In imaging mode, the plate scale is 0.15 arcsec/pixel and the field of view is 7.74 arcmin in diameter (3096 × 3096 unbinned pixels). We used binning in 2 × 2. The photometry was done with Bessel BVRI filters. In order to increase the signal-to-noise ratio for image analysis, we co-added the images obtained for each filter. To perform an absolute flux calibration of the comet images three field stars from APASS catalog for the photometric reference AAVSO (https://www.aavso.org/apass) were used. The seeing was stable during the night, with a mean value of 0.8″ FWHM. We processed data using the IDL software, which included basic bias subtraction, flat-field correction, and removal of cosmic ray traces. The twilight sky was used to provide flat-field corrections. The sky background level was estimated from a frame free from the cometary coma and bright stars. The combination of all images was done using the central isophote. For quantitative investigations of morphological features in cometary coma, the stars in the vicinity of the comet were removed by the two-dimensional approximation method.

**Table 1**
Log of the observations of comet 29P/Schwassmann-Wachmann 1.

| Date, U | $T_{exp}$,s [a] | Number of frames | Filters |
| --- | --- | --- | --- |
| 2016 August 12.192 | 300 | 8 | B |
| 2016 August 12.207 | 200 | 8 | V |
| 2016 August 12.240 | 100 | 5 | R |

### Coma morphology

Usually comet 29P presents outburst activity (Berman and Whipple, 1928a; Berman and Whipple, 1928b; Trigo-Rodríguez et al., 2010; Ivanova et al., 2016). Jean-Francois Soulier photographed yet another 29P eruption on July 28, 2016 (Soulier, 2018). R. Miles (private communication) has indicated to us that he also saw an outburst near that time, observing around July 27.75 with an outburst amplitude of 3.30 mag. Miles also communicated to us that he saw a smaller outburst of amplitude 0.45 mag from observing around August 16.50. Our observations were made during August 12, so in between these two observing times. As the behavior of comet 29P is unique, the outburst status on August 12, is uncertain.

For all data obtained using BVRI filters, we applied an enhancement technique to highlight low-contrast structures in the cometary coma. We used a rotational gradient method (Larson and Sekanina, 1984), division by azimuthal average and azimuthal renormalization filtering (Samarasinha and Larson, 2014; Martin et al., 2015). Before applying the filtering, we removed all field stars near the nucleus and coma in every image.

*Larson–Sekanina filter.* This is one of the most applied filters to study cometary morphology used for the first time by Larson and Sekanina (1984). After filtering, the resulting images lose all possible photometric information, but will reveal hidden variations of brightness inside the coma.

*Division by azimuthal average.* In the case of gas species and for non-spherically symmetric outflows, the azimuthally averaged profile will generate a more representative background. The division by azimuthal average is expressed by Samarasinha and Larson (2014) and Martin et al. (2015) as

$$\begin{cases} I_{out}(x, y) = \frac{I_{in}(x,y)}{I_{aziave}(\rho)} \\ I_{aziave}(\rho) = \frac{\sum_{ij} I_{in}(x,y)}{n_\rho} \\ \rho^2 = x^2 + y^2 = x_j^2 + y_j^2, \end{cases} \quad (1)$$

where $I_{out}(x,y)$ and $I_{in}(x,y)$ are, respectively, output and input pixel values, $x, y, x_j, y_j$ are rectangular coordinates centered at the nucleus, $\rho$ is cometocentric distance projected in the sky plane.

*Azimuthal renormalization.* The renormalization is carried out in such a way to guarantee that post enhancement the respective minimum and maximum pixel values are $i_{min}$ and $i_{max}$ for all $\rho$. Generally, $i_{min}$ is chosen to be zero. We use a clipping algorithm to exclude outlier pixel values due to bright stars, cosmic rays, dead pixels, etc. The azimuthal renormalization is expressed by Samarasinha and Larson (2014) and Martin et al. (2015) as

$$I_{out}(x, y) = i_{min} + (i_{max} - i_{min}) \frac{I_{in}(x, y) - I_{min}(x_i, y_j, \rho)}{I_{max}(x_i, y_j, \rho) - I_{min}(x_i, y_j, \rho)} \quad (2)$$

$I_{min}(x_i, y_j, \rho)$ and $I_{max}(x_i, y_j, \rho)$ are, respectively, the minimum and maximum pixel values for a given $\rho$ in the original image and $\rho, x, y, x_i,$ and $y_j$ are related by Eq. (1).

More detailed information about the different enhancement techniques available are presented in http://www.psi.edu/research/cometimen. Earlier this technique was used to pick out structures in several comets with good results (Manzini et al., 2007; Ivanova et al., 2009; Ivanova et al., 2016; Ivanova, 2017; Miles et al., 2016; Rosenbush et al., 2017). Fig. 1 shows unprocessed and processed images using various digital filters.

The available observations of comet 29P obtained by Roemer (1958) and the next apparitions of the comet showed that the structure of its cometary coma were non-isotropic and time-dependent. The comet demonstrates outburst activity and the coma is usually asymmetric in form (Miles et al., 2016).

In all processed images of Fig. 1, three morphological features are seen in all filters at approximately the same orientation. Structures appear as radial outflow of dust from the nucleus. If we compare the structures for different filters, we can see them changing shape and intensity. These changes are probably related to the size and velocity of the dust particles outflowing from nucleus surface. Based on the classification of structures presented in Farnham (2009), we define these features as a fan and jets. We note a fan visible in the South–West quadrant of Fig. 1 pointing toward the Sun. Three jets (marked as 1, 2, 3 in Fig. 1) emanate from the nucleus, but they are narrower and more sharply defined than the fan. Also we can see in Fig. 1, that the morphology of structures in different filters are non identical. For example in Filter R, these differences are significant. Probably we see changes of reflecting dust properties with wavelength. These outflows appear to be related active areas on the nucleus surface (Senay and Jewitt, 1994), but in reality, according to the Rosetta observations of comet 67P/Churyumov-Gerasimenko, there were no such areas found having very strong and long-lasting activity, which could be associated with jets observable from the Earth (Lin et al., 2016). Analysis of ground-based long-term monitoring of comet 29P (Gunnarsson, 2003; Miles et al., 2016) has shown that such comet asymmetry and fan-shaped comae are consistent with an initial expansion phase whereby dust grains are accelerated by gas originating from the sublimation of volatile ices. Visible-wavelength and radio observations confirmed that the comet contains $CO^+$ and $N_2^+$ (Cochran and Cochran, 1991; Korsun et al., 2008; Ivanova et al., 2016; 2018) both of which may lead to its unique behavior since their presence indicates the presence of volatiles species that can exist in the comet nuclei in the form of ices, as well as in the form of gaseous inclusions in amorphous water ice cells.

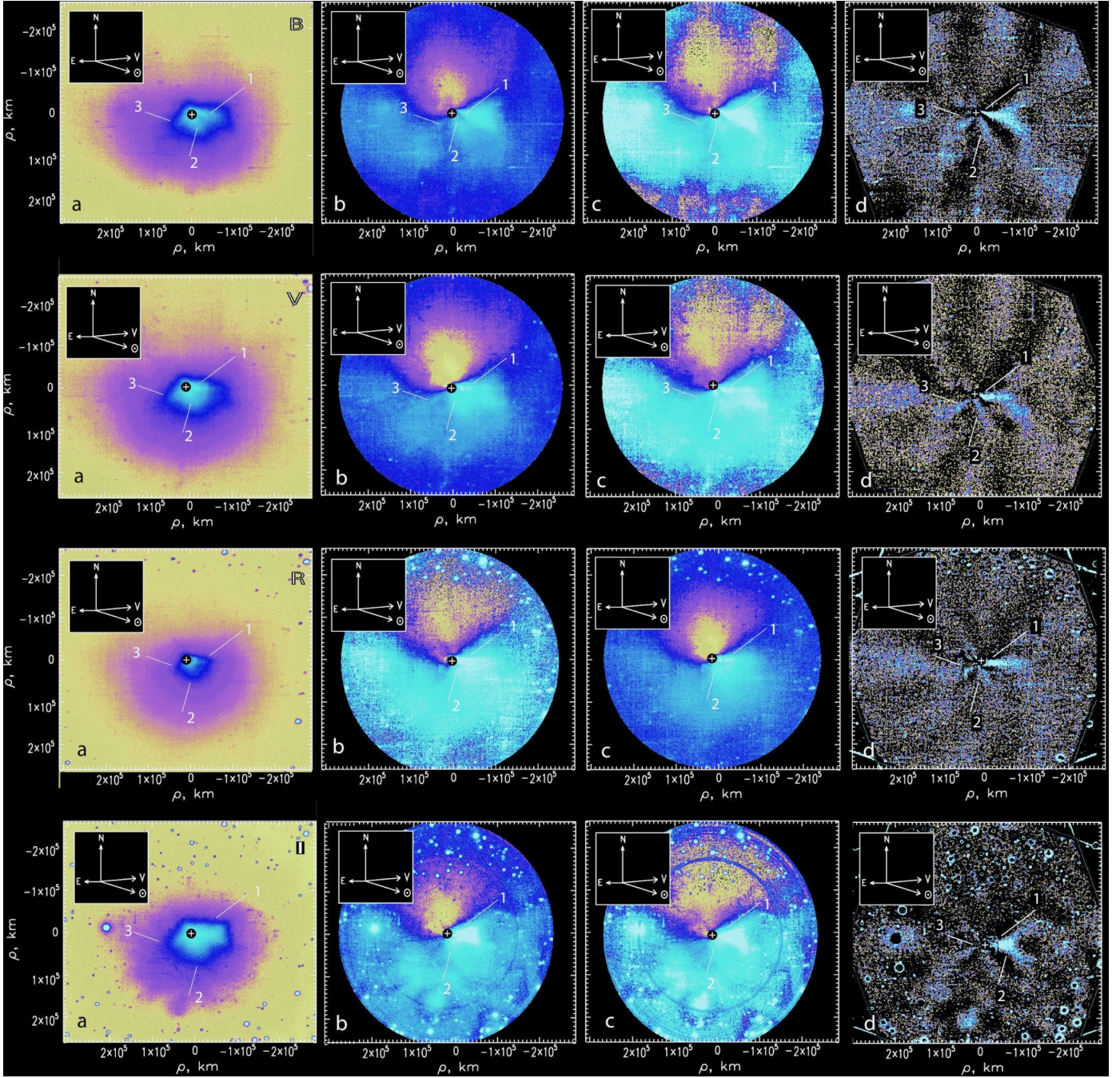

**Fig. 1.** Intensity images of comet 29P were obtained with the broad-band filters treated with digital filters: (a) direct image; (b) azimuthally average filter; (c) azimuthal renormalization filter; and the (d) Larson–Sekanina filter. North, East, sunward, and velocity vector directions are indicated. The white circle with cross inside correspond to photocenter of comet. The big circles in the fourth row, panel (b) and (c) are artifact from using of digital filters.

Hill et al. (2001) argue that comets formed early in nebular history should be rich in CO, $CO_2$, $N_2$, and amorphous dust. CO sublimation begins at about 25 K (at large distances), while CO gas release from water ice cells begins at temperatures more than 100 K (Prialnik et al., 1995), i.e. at distances closer to the Sun. It should be noted that the orbital distance of 29P is within the 100 K boundary, but CO sublimation can depend a lot on how heat flows into the interior of the nucleus.

Observations of $CO^+$ and $N_2^+$ are important because they are tracers of their parent molecules. CO is difficult to photoionize at large heliocentric distances (more than 5 AU), so detecting $CO^+$ implies probably a large amount of CO being released in the coma. Senay and Jewitt (1994) reported that CO emission first detecting at millimeter wavelengths in comet 29P in amounts significantly higher than the dust production rate. Carbon monoxide was detected in comet 29P again with IRAM by Crovisier et al. (1995). Recent ground-based infrared observations of the comet (Paganini et al., 2013) have confirmed that CO can be the main driver of the 29P cometary activity.

Other highly volatile components, such as $CO_2$ and $N_2$ may contribute to distant activity of the comet 29P. But as showed in Cochran et al. (2000) and Iro et al. (2003) $N_2$ does not play a significant role in activity of comet at large heliocentric distances. Also in the discussion presented by Womack et al. (2017) the authors conclude that all available data indicate that $CO_2$ is not likely responsible for most of the observed activity in 29P.

## 4. Analysis of photometrical observations

The magnitude of the coma is determined as following:

$$m_c(\lambda) = -2.5 \cdot lg\left[\frac{I_c(\lambda)}{I_s(\lambda)}\right] + m_{st} - 2.5 \cdot lgP(\lambda) \cdot \Delta M, \quad (3)$$

where $m_{st}$ and $m_c$ are, respectively, the integrated magnitude of the standard and comet, $I_c$ and $I_s$ are the measured fluxes of the comet and star, respectively; P is the sky transparency which depends on the wavelength $\lambda$; and $\Delta M$ is the difference between the cometary and the stellar air masses. As we applied the field stars for calibration, the sky transparency is not considered. The apertures used for the photometry for star and comet were the same.

Using the calculated magnitude of the coma, we estimate the color indexes and create color maps (Fig. 2) to analyse color variation globally over the comet and its local structures. The errors in the magnitude measurements vary between 0.01 and 0.03 m.

Fig. 2 shows color maps (B–V), (B–R) and (B–I) of the comet. As well as in the filtered images, we can see clearly the structure in the color map.

In the processed color B–I image (in Fig. 2), jet 1 is seen.

To investigate the color trends along different directions, we have taken cuts through the photometric center of the comets. The cuts along the solar direction (Sun), perpendicularly to the solar direction, and along jet are presented in Fig. 3. The near-nucleus area of the comet (up to approximately 10000 km) has a very red color, on average, 0.8, 1.4 and 2.4 m for B–V, B–R and B–I, respectively. In general, the color of the coma (B–V and B–R) becomes bluer with increasing distance from the nucleus suggesting an evolution of the dust particles. One can see that color of structure presented on color map B–I is different than other coma. The color of the jet is noticeably bluer than the surrounding coma. This may indicate that the dust contains smaller particles or/and has a different (ices?) composition.

We calculate the coma magnitude in a circular aperture centered on optocenter (Tables 2, 3) and color indexes.

Our analysis shows that the changing colors of the coma are consistent with changing dust properties during an outburst. Walker (1959), Kiselev and Chernova (1979), and Miles et al. (2016) have noted that the changing dust colors during and after outbursts varied from red to blue.

We compute the Af$\rho$ (A'Hearn, 1984) parameter, for an analysis of dust activity (Table 4) and absolute magnitudes of the comet. The reduced magnitude can be computed as

$$H(1, 0, 0) = R - 5 \cdot \log(r \cdot \Delta) - \beta \cdot \alpha, \quad (4)$$

where $r$ is the heliocentric distance (AU), $\Delta$ is the geocentric distance (AU), $\alpha$ and $\beta$ are the phase angle during the observations (°) and phase coefficient for dust (mag/deg) respectively. Lowry et al. (2003) assumed that $\beta$ would be the same for dust and nuclei, so they just used 0.035 mag/deg value for the active comets in their paper. Ciarniello et al. (2015) report coefficients that are for nuclei and make no statements about dust. It looks like the numbers in Ciarniello's Table 3 show a median or average value higher than 0.04. We use $\beta = 0.4$ mag/deg as a good compromise between the Lowry et al. (2003) report and the Ciarniello et al. (2015) report. Af$\rho$ can be derived from the calculated photometric magnitude (Mazzotta Epifani et al., 2009)

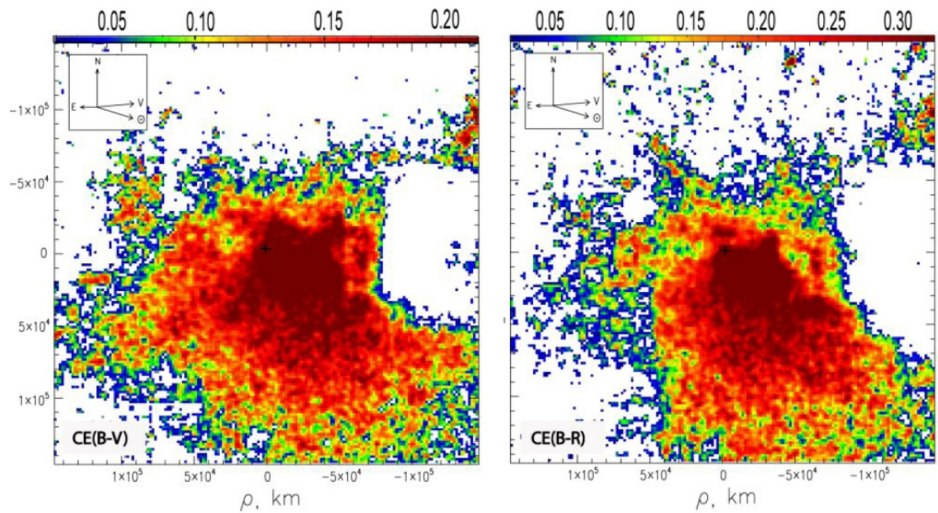

**Fig. 2.** Color maps of comet 29P derived on August 12, 2016. The maps are painted according to color indexes in magnitudes, as indicated in the bar at the top of the images. Arrows point in the direction to the Sun (☉), North (N), East (E), and velocity vector of the comet as seen on the observers sky plane (V). Map B–V is shown at left. Map B–R s shown at the right. (For interpretation of the references to color in this figure legend, the reader is referred to the web version of this article.)

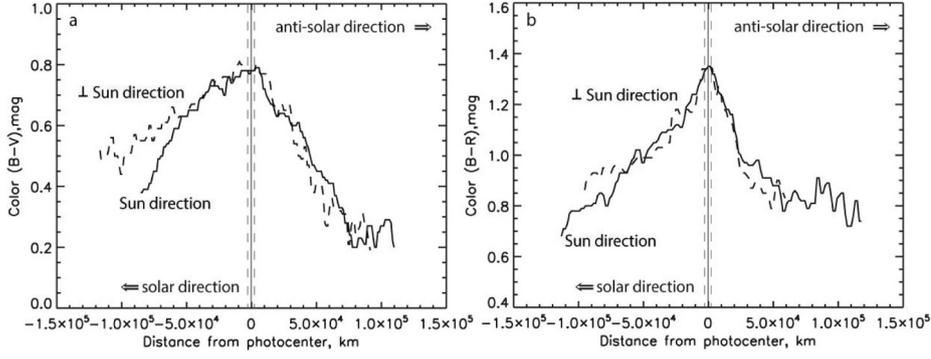

**Fig. 3.** Scans across the actual color map B–V (a) and B–R (b) of comet 29P along the solar (denoted as Sun direction) and perpendicularly to the solar anti-solar direction (denoted as ⊥Sun direction), along the jet (Jet). Positive distance is in the antisolar direction, and negative distance is in the solar direction. Zero point is at the photometric centre of the comet. Vertical dashed lines show the size of the seeing disk during the observations. (For interpretation of the references to color in this figure legend, the reader is referred to the web version of this article.)

**Table 2**
Total magnitudes of the comet 29P.

| ρ(arcsec) | mB | mV | mR | mI |
|---|---|---|---|---|
| 1.1 | 18.32 ± 0.03 | 17.54 ± 0.02 | 16.97 ± 0.01 | 15.91 ± 0.03 |
| 2.2 | 17.56 ± 0.03 | 16.76 ± 0.02 | 16.21 ± 0.01 | 15.09 ± 0.03 |
| 3.3 | 17.06 ± 0.03 | 16.29 ± 0.02 | 15.74 ± 0.01 | 14.64 ± 0.03 |
| 6.6 | 16.26 ± 0.03 | 15.48 ± 0.02 | 14.97 ± 0.01 | 13.87 ± 0.03 |

$$Af\rho = \frac{4 \cdot r_h^2 \cdot \Delta^2 \cdot 10^{0.4 \cdot (m_{sun} - m_c)}}{\rho}, \quad (5)$$

where $m_{sun}$ is the Sun magnitude and $m_c$ is the magnitude (in the same filter) in an aperture of radius ρ. We used the following values for Sun magnitudes: −26.09 (B), −26.74 (V), −27.26 (R) and −27.59 (I). For 29P the values of Afρ change with cometocentric distance, and the calculated average value of Afρ lies in the interval from 1.1″ to 6.6″, corresponding to cometocentric distances from 4,001 to 24,010 km, respectively.

Results suggest that 29P was active during our observations. The Afρ values are higher than our previous results for the comet, obtained in 2012 (Ivanova et al., 2016). The Afρ values of comet 29P are varying with time and can change from hundreds to thousands of centimeters in a few days (Hosek et al., 2013; Trigo-Rodríguez et al., 2010). The maximum Afρ values (56,000 cm) in the R band were found in the comet during its outbursts in 2010 (Trigo-Rodríguez et al., 2010).

For an analysis of the observation, it is important to estimate the relative contribution of the gaseous emission into the measured flux. In order to do this, we estimate the level of gas emission contamination in broadband images. One can compute such a ratio from the 29P spectrum reported by Ivanova et al. (2018) for the same epoch with our present BVRI photometric observations. Fig. 4 shows the continuum and the emission spectra of 29P (data are adapted from Ivanova et al. 2018) as well as the transmittance curves of the B and V filters.

Using the data presented in Fig. 4, we computed the relative contribution of the pure gaseous emission into the total signal within the B and V filters. We found that in the former case the gaseous emission (there are numerous lines of $CO^+$ as well as the $N_2^+(0, 0)$ line of the ($B^2\Sigma$-$X^2\Sigma$) system) contributes about 2.5% of the total flux; whereas, in the latter case, it is of only 1%. Although the spectrograph used by Ivanova et al. (2018) does not allow to do similar analysis as was done in R and I filters, our computations unambiguously reveal that the gaseous emission is only a minor contributor to the total flux from 29P and, therefore, it could be omitted in the analysis. Also there are simply not as many strong gas emission bands in for R and I bands. Thus the gaseous emission is only a minor contributor to the total flux from 29P and, therefore, it could be omitted in the analysis.

In addition to the color index, the color in a comet is also often characterized with the so-called color slope S′ (e.g., A'Hearn, 1984; Lamy, 2009; Ivanova, 2017) that is defined as follows:

$$S' = \frac{10^{0.4\Delta m} - 1}{10^{0.4\Delta m} + 1} \times \frac{20}{\lambda_2 - \lambda_1}. \quad (6)$$

Here Δm is the true color index of the comet, i.e., the observed color reduced by the color of the Sun, and $\lambda_1$ and $\lambda_2$ stand for the effective wavelengths (μm) of the used filters. In Table 5 we present the color slope S′ measured in percent per 0.1 μm in comet 29P at different aperture sizes and pairs of filters. Note that the color slope is utilized in studies of cometary comae as well as in cometary nuclei.

A significant peculiarity of the color slope is that it is independent of the specific characteristics of the chosen filters. Indeed, as one can see in Eq. (6) the color slope is normalized to the difference in the filters wavelengths and, as a consequence, this makes possible a comparative analysis of the color slope inferred with different photometric systems. Furthermore, in some works, the color slope formally obtained in one waveband is extrapolated toward a significantly different waveband (e.g., Schambeau, 2015). The latter assumption, however, should be considered with caution as it has no strong physical basis that also was noticed in Schambeau (2015). In particular, our current measurements summarized in Table 5 reveal a noticeable dependence of the color slope in the 29P coma on waveband. Namely, in the pair of filters R–I the color slope appears to be twice of what is in the pair B–V.

**Table 3**
Color of the comet 29P.

| | $m_c$–$m_{sun}$ | | | | | |
|---|---|---|---|---|---|---|
| ρ(arcsec) | B–V[a] | B–R | B–I | V–R | V–I | R–I |
| 1.1 | 0.14 ± 0.05 | 0.36 ± 0.05 | 1.08 ± 0.05 | 0.22 ± 0.05 | 0.94 ± 0.03 | 0.73 ± 0.03 |
| 2.2 | 0.16 ± 0.05 | 0.36 ± 0.05 | 1.14 ± 0.05 | 0.20 ± 0.05 | 0.98 ± 0.03 | 0.79 ± 0.03 |
| 3.3 | 0.13 ± 0.05 | 0.33 ± 0.05 | 1.09 ± 0.05 | 0.20 ± 0.05 | 0.96 ± 0.03 | 0.77 ± 0.03 |
| 6.6 | 0.14 ± 0.05 | 0.30 ± 0.05 | 1.06 ± 0.05 | 0.16 ± 0.05 | 0.92 ± 0.03 | 0.77 ± 0.03 |

[a] Color index of the Sun from Holmberg et al. (2006) (B–V = 0.64 ± 0.04; B–R = 0.99 ± 0.04; B–I = 1.33 ± 0.04; V–R = 0.35 ± 0.04; V–I = 0.69 ± 0.014; R–I = 0.33 ± 0.008)

**Table 4**
Absolute magnitudes of the comet and Afρ values (expressed in cm) versus cometocentric distance.

| ρ(arcsec) | mB(1,0,0) | mV(1,0,0) | mR(1,0,0) | mR(1,0,0) | AfρB | AfρV | AfρR | AfρI |
|---|---|---|---|---|---|---|---|---|
| 1.1 | 10.75 ± 0.03 | 9.97 ± 0.02 | 9.40 ± 0.01 | 8.34 ± 0.03 | 3393 ± 93 | 3825 ± 70 | 4005 ± 76 | 7846 ± 220 |
| 2.2 | 9.99 ± 0.03 | 9.19 ± 0.02 | 8.64 ± 0.01 | 7.52 ± 0.03 | 3417 ± 94 | 3923 ± 72 | 4033 ± 77 | 8349 ± 230 |
| 3.3 | 9.49 ± 0.03 | 8.72 ± 0.02 | 8.17 ± 0.01 | 7.07 ± 0.03 | 3610 ± 99 | 4032 ± 74 | 4145 ± 78 | 8425 ± 230 |
| 6.6 | 8.69 ± 0.03 | 7.91 ± 0.02 | 7.40 ± 0.01 | 6.30 ± 0.03 | 3771 ± 104 | 4251 ± 78 | 4212 ± 78 | 8561 ± 240 |

In order to analyze our measurements quantitatively, we consider the model of *agglomerated debris particles*. Such particles have highly irregular aggregate morphology (see ten samples in Fig. 5) with packing density of their constituent materials being in accordance with what was found *in situ* in micron-sized cometary and in interplanetary dust particles (Zubko, 2016). It is important emphasize that the agglomerated debris particles have a proven capability to reproduce quantitatively the photopolarimetric observations of various comets. For instance, they satisfactorily reproduce spatial variations of the linear polarization degree often observed in comets (e.g., Zubko, 2015; and references therein), phase function of brightness and polarization simultaneously measured in a comet over a wide range of phase angles (Zubko, 2014), dispersion of positive linear polarization in different comets (Zubko, 2016), and both photometric and polarimetric colors of various comets (Zubko, 2011; Zubko, 2014; Zubko, 2015; Zubko, 2016; Ivanova, 2017).

We computed the light scattering by agglomerated debris particles with the discrete dipole approximation (DDA), a flexible approach that places minimum restrictions on the shape of target particles. We refer to our previous works, (Zubko, 2014; 2015; 2016; Ivanova, 2017), for an exhaustive discussion of various practical aspects in the DDA modeling of photometric (and polarimetric) color of dust in comets; below we present just a short introduction into this problem.

Light scattering by micron-sized particles is governed by the ratio of their size to the wavelength of the incident light λ. In the literature, it is quantified in terms of the *size parameter* $x = 2\pi r/\lambda$, where r stands for radius of the particle (e.g., Bohren and Huffman, 1983). Another important characteristic of target particles is the *refractive index m* that characterizes the ability of the constituent material to absorb and scatter light. The wavelength dependence of the light-scattering response can be caused by changes of the size parameter x and refractive index m with λ.

Cometary dust is long known to be polydisperse in size with a significant number of submicron and micron-sized particles (e.g., Mazets, 1986; Price, 2010). However, the relative contribution of a specific size into the total light-scattering response is determined not only by size distribution. Another important factor is the *scattering efficiency* $Q_{sca}$, the characteristic describing the ratio of flux of the total scattered electromagnetic energy (i.e., integrated over a circumscribing sphere around the particle) to the flux of incident electromagnetic energy that flows through the geometric cross section of that particle (e.g., Bohren and Huffman, 1983). At $x < 1$, $Q_{sca}$ is substantially less than 1, implying that only a minor part of the incident electromagnetic radiation interacts with a target particle. $Q_{sca}$ grows quickly with size parameter x. In irregularly shaped particles, $Q_{sca}$ reaches its maximum values of approximately 1.5–3 at $x = 6–8$ (Zubko, 2013). In a perfect sphere this phenomenon appears much stronger; $Q_{sca}$ can be as high as ∼ 10 that occurs at $x \approx 3–4$ (Bohren and Huffman, 1983). In other words, particles with size comparable to wavelength are very efficient at light scattering and they scatter more electromagnetic energy than flows through their geometric cross section. It is significant that a further increase of x leads to a systematic decay of $Q_{sca}$ until the size of particle becomes sufficient for geometric optics approximation. Such a profile of $Q_{sca}$ vs. x dampens the contribution of small particles ($x < 1$) and relatively large particles ($x > 20$), simultaneously, this enhances the role of particles whose size is comparable to the wavelength ($x \sim 10$). In addition to this dampening of $Q_{sca}$, the light scattering response of larger particles is greatly diminished due to the power-law size distributions consistently measured in cometary comae. As demonstrated by Zubko (2013), at phase angle $\alpha < 30°$, light scattering by irregularly shaped particles is mainly caused by size parameter $x \leq 15$. In the present study, we consider particles with values of x that are at least 1.5 times larger than this.

**Table 5**
The color slope S′ [in % per 0.1 μm] vs. aperture diameter in comet 29P/Schwassmann-Wachmann 1 for various pairs of filters.

| ρ(arcsec) | ρ(km) | B–V | V–R | B–R | R–I |
|---|---|---|---|---|---|
| 1.1 | 4001 | 12.8 ± 4.6 | 21.1 ± 4.8 | 16.7 ± 2.3 | 27.3 ± 1.0 |
| 2.2 | 8003 | 14.6 ± 4.6 | 19.2 ± 4.8 | 16.7 ± 2.3 | 29.3 ± 1.0 |
| 3.3 | 12005 | 11.9 ± 4.6 | 19.2 ± 4.8 | 15.4 ± 2.3 | 28.6 ± 1.0 |
| 6.6 | 24010 | 12.8 ± 4.6 | 15.4 ± 4.8 | 14.0 ± 2.3 | 28.6 ± 1.0 |

We utilize computational results previously obtained for agglomerated debris particles at two different sets of x. The first set embraces the size parameter from $x = 1$ up to 32, with only few exceptions for the largest value of x caused by dependence of the DDA convergence on refractive index m. The second set of x is chosen to correspond to a 1.432-times longer wavelength as compared to the wavelength in the first set of x. As a consequence, the second set of size parameter ranges from x ≈ 0.7 up to x ≈ 22.3. Based on these two sets, one can compute the color slope S′. It is important that the obtained color slope S′ can be applied equally to the analysis of observations conducted with various pairs of filters. The pairs only should have a similar difference in their effective wavelengths, i.e., being close to a factor of 1.432. In our set of BVRI filters, this requirement is met in two pairs of filters, B–R and R–I, which have ratio of their wavelengths of 1.454 and 1.378, respectively. They are very close to the factor used in the modeling, with a deviation less than 4%.

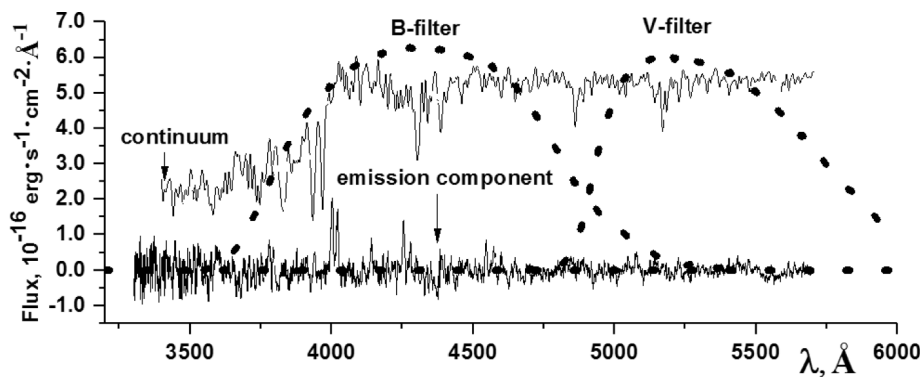

**Fig. 4.** Continuum, the emission spectrum, and the filters transmission curves for the comet 29P/Schwassmann-Wachmann 1 (this one is based on the data from Ivanova et al., 2018). The filters curves were taken from http://www.ctio.noao.edu/soar/content/filters-available-soar.

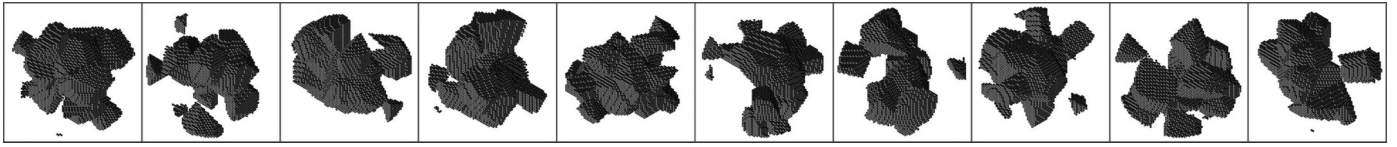

**Fig. 5.** Ten examples of the agglomerated debris particles.

The color slope S′ in comet 29P measured with the pairs of filters B–R and R–I are given in two columns on the right in Table 5. As one can see, it is quite difficult to quantify a systematic change in S′ with size of the aperture. If it exists it should be smaller than the uncertainty in our measurements. Therefore, in what follows, we analyze the aperture-averaged color slopes corresponding to the largest available aperture that are given on the bottom in Table 5.

We first investigate the case of a wavelength-independent refractive index m. While m is invariant, the size parameter x changes with wavelength λ, which solely produces photometric color. This type of wavelength dependence was found to reproduce the color of dust in some comets (e.g., Zubko, 2014; Ivanova, 2017) or, at least, in some part of cometary coma (e.g., Zubko, 2015). Fig. 6 demonstrates the color slope in the agglomerated debris particles at phase angle α = 5° and at 35 different wavelength-independent refractive indices; see the legend for specific values of the refractive indices and Zubko (2015) and Ivanova (2017) for a discussion of their relevance to cometary species. The color slope is shown as a function of the index n in a power-law size distribution $r^{-n}$. The range of the considered power index n embraces the in situ findings on micron-sized fraction of dust in comets, e.g., n = 1.5–3 (see, e.g., Mazets, 1986; Price, 2010).

What immediately emerges from Fig. 6 is that none of the 35 materials can reproduce the color slope measured in comet 29P; even with the pair of filters B–R. It may imply that the materials with wavelength-independent refractive index do not appear in comet 29P in considerable amounts on the epoch of our observation; however, this does not necessarily hold on other epochs though. However, it also means that we need to consider a more sophisticated model of dust in this comet.

In Fig. 7 we consider agglomerated debris particles consisting of a material with wavelength-dependent refractive index. In all the cases, real part of the refractive index Re(m) takes on the same value at both wavelengths; whereas, the imaginary part Im(m) at shorter wavelength is assumed to be somewhat greater compared to what it is at longer wavelength. As discussed, for instance, in Zubko (2015), this is a common trend in various cometary species. Three panels in Fig. 7 show from top to bottom the increment of ΔIm(m) = 0.01, 0.02, and 0.03, respectively. Each curve in Fig. 7 is characterized with two refractive indices (see the legend). The one denoted with the symbol S corresponds to a shorter wavelength and another one marked with the symbol L, to a longer wavelength. When a curve is considered in application to data obtained with the pair of filters B–R, symbol S should be attributed to the B filter and symbol L to the R filter; whereas, in the pair R–I, S corresponds to the R filter and L to the I filter. In addition to computational results, we show also the color slope measured in 29P with two horizontal lines.

As one can see, the modeling results in Fig. 7 appear in much better accordance with observations of comet 29P as compared to Fig. 6. At ΔIm(m) = 0.02 and 0.03 (middle and bottom panels, correspondingly), one can fit the color slopes in both pairs of filters. However, the smallest increment, ΔIm(m) = 0.01, fails to reproduce the R–I color slope.

A self-consistent fit to the color slopes in both pairs of filters places two important restrictions on model particles. First, both color slopes must correspond to the same power index n in the size distribution. Second, refractive index of the particle material in the R filter must be the same in modeling of the B–R color slope and of the R–I color slope. Analysis of Fig. 7 reveals only one such solutions that simultaneously meets these requirements. It occurs at the power index n ≈ 2.55. In Fig. 7 it is marked with blue solid circles in the middle and in the bottom panels; whereas, the radius of the circle is equal to the error bars in the colorimetric measurements. In this scenario, the B–R color slope is approximately reproduced with the refractive index m = 1.6 + 0.07i in the B filter and m = 1.6 + 0.05i in the R filter; whereas, the R–I color slope is fitted at m = 1.6 + 0.05i in the R filter and m = 1.6 + 0.02i in the I filter.

It is important to point out that the power index n ≈ 2.55 appears in good quantitative agreement with the in situ findings in comets 1P/Halley and 81P/Wild 2 (Mazets, 1986; Price, 2010) as well as with the modeling results of the ground-based polarimetric measurements of various comets (e.g., Zubko, 2016).

Our modeling reveals that the wavelength dependence of the imaginary part of refractive index Im(m) is as follows. At λ = 0.4326 μm, Im(m) = 0.07; at λ = 0.6289 μm, Im(m) = 0.05; and, at = 0.8665 m, Im(m) = 0.02. This very much resembles what was experimentally detected in Mg-Fe pyroxene glasses (see Table 5 of Dorschner, 1995). From the in situ study of comet 1P/Halley, silicates are predominantly Mg-rich and Fe-poor (e.g., Fomenkova, 1992), but small amounts of Mg-Fe silicates were detected in situ in comet 81P/Wild 2 (Zolensky, 2007). They also may fit mid-IR spectra (Wooden, 2008) and polarization of comets (Zubko, 2015; Ivanova, 2017). In addition, Fe-Mg silicates have been previously inferred in 29P (Schambeau, 2015).

Finally, we would like to discuss a shortcoming of the present analysis that arises from the limited amount of observational data. Namely, we managed to organize the observation of comet 29P only on a sole epoch, 2016-08-12. As a consequence, our conclusion on chemical composition of its coma must be considered only with regard to that epoch. It is known that the color in comets could be subject to fast and dramatic variations (e.g., Ivanova, 2017) that reflects significant changes of their coma dust population. Such variations have to be a fortiori expected in comet 29P whose nucleus is currently disintegrating; and, therefore, dust population of its coma can be unlikely characterized exhaustively with a sole observation. Nevertheless, our simultaneous analysis of the color slope inferred with two sets of filters put quite strong constraints on the 29P coma composition, suggesting its primarily Mg-Fe-silicate composition. To date, we accomplished quantitative analyses of various comets (e.g., Zubko, 2014; Zubko, 2015; Zubko, 2016; Ivanova, 2017). Prior to this work, we found a chemically homogeneous coma only in one comet, 17P/Holmes (Zubko, 2011); whereas, in all other cases, we retrieve at least two different refractory components of coma. The retrieval of a single composition of particles in 29P inferred in this study does not immediately imply that this is a persistent feature of the given comet. This matter requires further investigation of comet 29P.

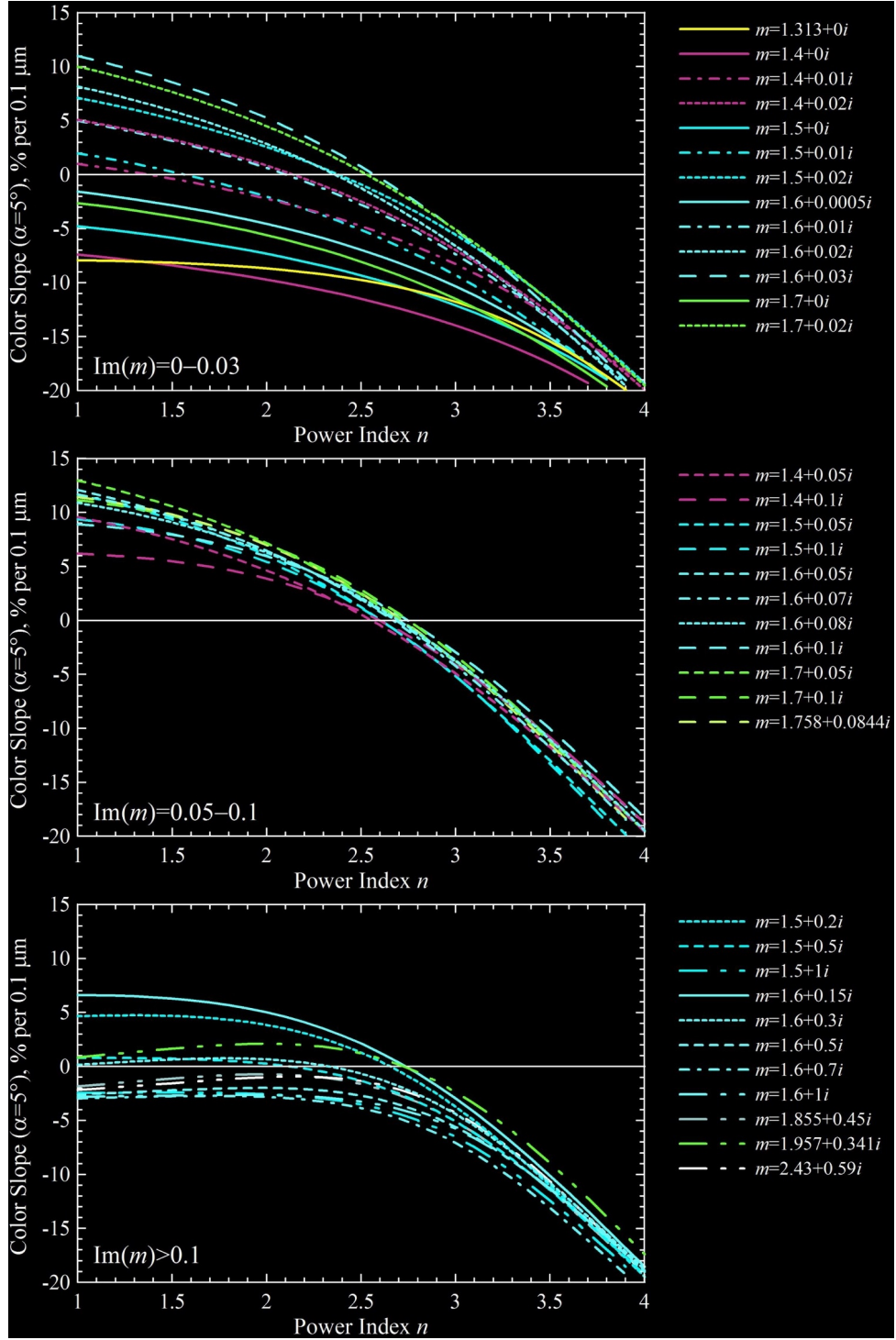

**Fig. 6.** Color slope S′ at phase angle $\alpha=5°$ in the agglomerated debris particles as a function of the index n in their power-law size distribution $r^n$. The case of wavelength-independent refractive index m. (For interpretation of the references to color in this figure legend, the reader is referred to the web version of this article.)

**Summary**

The main results provided by our observations can be summarized as follows:

I   Comet 29P was active during the night of observations on August 12, 2017;
II  Three radial features are identified in the comet 29P coma from the image-processing algorithm;
III The level of gas emission contamination in filtered images was small, 2.5%;
IV  The calculated dust production Afρ in comet 29P was typical for outburst activity of this object: from 3,393 ± 93 cm to 8,561 ± 236 cm for different filters;
V   The model of agglomerated debris particles is used for simultaneous analysis of the color slope in the B–R and R–I pairs of filters. The chemical composition of the 29P dust particles is consistent with Fe–Mg silicates and obey a power-law size distribution with index n ≈ 2.55. This conclusion is consistent with previous thermal-emission studies of 29P/Schwassmann-Wachmann 1. Although there is always a chance that the problem of modeling of the 29P observations has a non-unique solution, it also is worth noting that we did not manage to find an alternative solution using an existing extended database;

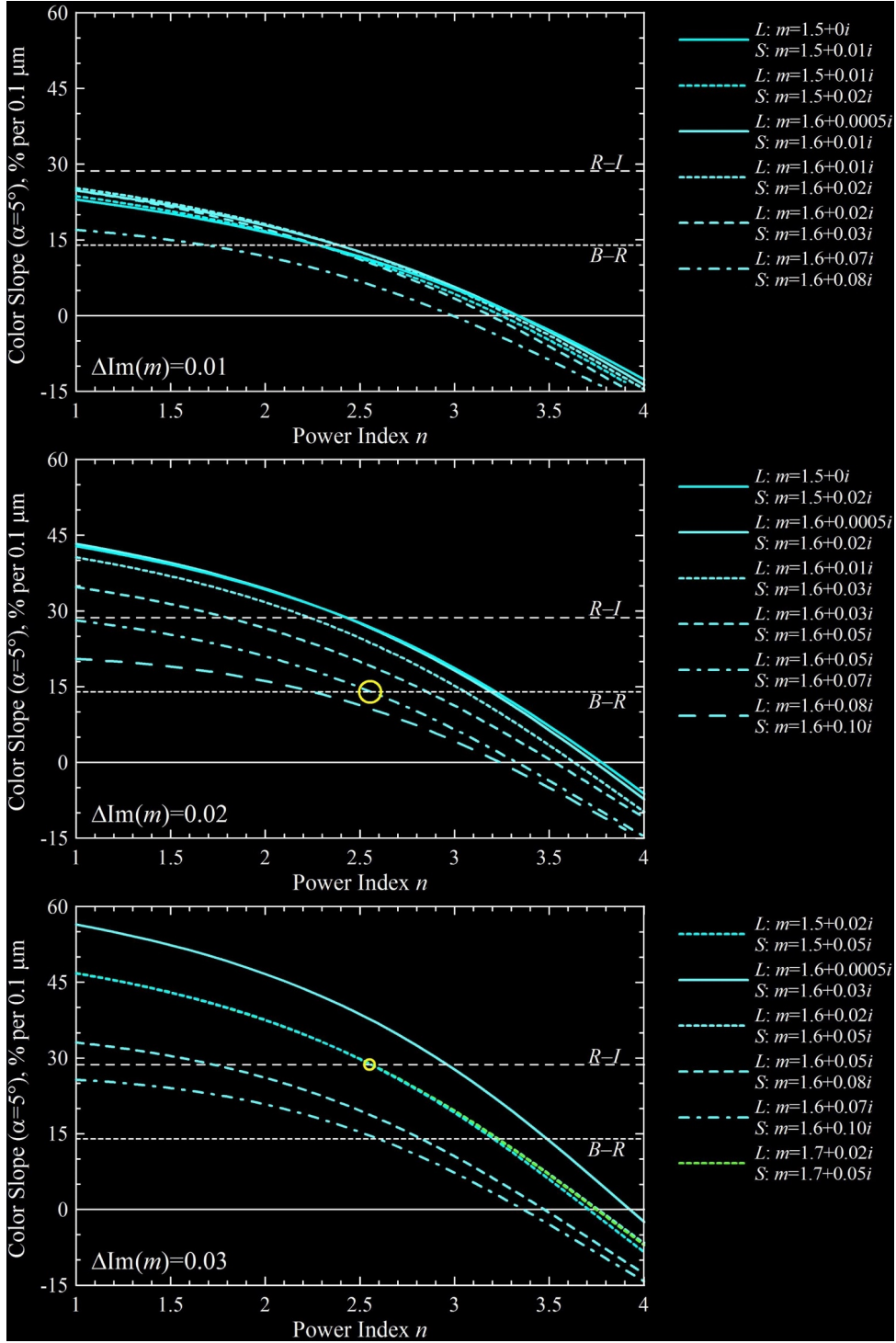

**Fig. 7.** Color slope S′ at phase angle $\alpha = 5°$ in the agglomerated debris particles as a function of the index n in their power-law size distribution $r^n$. The imaginary part of refractive index Im(m) decreases while the wavelength λ grows. From top to bottom the increment of ΔIm(m) is 0.01, 0.02, and 0.03, respectively. The dashed horizontal lines demonstrate the color slope in comet 29P. See text for more detailed discussion. (For interpretation of the references to color in this figure legend, the reader is referred to the web version of this article.)

## Acknowledgments

This research is based on observations obtained at the Southern Astrophysical Research (SOAR) telescope, which is a joint project of the Ministério da Ciência, Tecnologia e Inovação (MCTI) da República Federativa do Brasil, the U.S. National Optical Astronomy Observatory (NOAO), the University of North Carolina at Chapel Hill (UNC), and Michigan State University (MSU). O. Ivanova thanks the SASPRO Programme, the People Programme (Marie Curie Actions) European Union's Seventh Framework Programme under REA grant agreement No. 609427, and the Slovak Academy of Sciences (grant Vega 2/0023/

18). I. Luk'yanyk research is supported by the project 16BF023-02 of the Taras Shevchenko National University of Kyiv and the Slovak Academy of Sciences (grant Vega 2/0023/18). Also authors thank two anonymous referees for their constructive reviews.